\documentclass[10pt,a4paper]{article}
\usepackage[utf8]{inputenc}
\usepackage{multicol}
\usepackage{amsmath}
\usepackage{amsfonts}
\usepackage{amssymb}
\usepackage{graphicx}
\usepackage{hyperref}
\usepackage{authblk}

\newcommand{\jpsi}{J/\Psi}
\begin{document}
\title{Exclusive diffractive vector meson production:
A comparison between the dipole model and the Leading Twist Shadowing approach}
\author[1]{Shashank Anand\footnote{Shashank.Anand@physik.uni-muenchen.de}}
\author[2]{Tobias Toll \footnote{tobiastoll@iitd.ac.in}}
\affil[1]{Department of Physics, Ludwig-Maximilians-University \\Schellingstrasse 4, 80799 Munich, Germany} 
\affil[2]{Physics Department, Indian Institute of Technology Delhi\\
~~~Hauz Khas, New Delhi 110016, India}
\maketitle

\begin{abstract}
With the imminent construction of an electron-ion collider in the USA, both saturation physics and nuclear shadowing physics will be important for the same processes for the first time. In particular for exclusive production of vector mesons, such as the $\jpsi$. In this paper we investigate the underlying assumptions commonly made when phenomenologically investigating this process from a shadowing and saturation perspective respectively. We use the bSat model which is commonly used to describe saturation physics, and a calculation by M.G. Ryskin which is commonly used when describing nuclear shadowing at leading twist. It is expected that the bSat model is equivalent to Ryskin's result in the hard scattering and non-relativistic limits. By explicitly taking these limits we show that the bSat model does indeed become equivalent to Ryskin's result. However, two more approximations are needed for this result. Firstly, the factorisation and renormalisation scales in the bSat model has to be independent of the dipole radius, and secondly, we need to omit a term in the overlap between the vector meson and virtual photon wave functions. We show that for the typical dipole radius of the $\jpsi$, the different approximations off-set each other and the bSat model agrees very well with Ryskin's calculation in hard-scattering and non-relativistic limits.
\end{abstract}

\newpage
\section{Introduction}
With the advent of an electron-ion collider \cite{eic} in the USA, the first ever
high energy high luminosity electron-nucleus collider experiment, phenomena of gluon saturation as well as of nuclear
shadowing will become important for the same processes for the first time. 
One process that especially gives overlap between the two kinds of phenomena is that of exclusive diffractive production of vector mesons, such as the $\jpsi$ meson.

In the process $e+p\rightarrow e+J/\Psi+p$, the differential cross-section is given by:
\begin{eqnarray}
	\frac{{\rm d}\sigma^{\gamma^*p\rightarrow J/\Psi p}}{{\rm d} t}=
	\frac{1}{16\pi}\left|\mathcal{A}^{\gamma^*p\rightarrow J/\Psi p}\right|^2
\end{eqnarray}
This is a diffractive process which at small momentum fractions $x$ can be seen as mediated by
a pair of virtual gluons. Here $\mathcal{A}^{\gamma^*p\rightarrow J/\Psi p}$ is the amplitude of the process. 

A widely used approach in the leading twist shadowing community (see e.g. \cite{Frankfurt:2011cs} and \cite{Baltz:2007kq}) for the amplitude was calculated in \cite{lts} by M.G. Ryskin:
\begin{eqnarray}
	\mathcal{A}_T^{\gamma^*p\rightarrow J/\Psi p}=i4\pi^2\sqrt{\frac{\Gamma_{ee}^{J/\Psi}M_{J/\Psi}^3}{3\alpha_{\rm EM}}}\alpha_s(\mu^2)xg(x, \mu^2) F_N^{2G}(t)\frac{2\mu^2+t}{(2\mu^2)^3}
\label{eq:ARyskin}
\end{eqnarray}
where $\Gamma_{ee}^{J/\Psi}$ and $M_{J/\Psi}$ are the $J/\Psi$ decay width and mass respectively, $F_N^{2G}(t)$ is the proton form-factor, which is a Fourier transform of its thickness in impact-parameter space. The strong coupling $\alpha_s$ and the gluon density $g$ are taken at a scale $\mu^2=(Q^2+M_{J/\Psi}^2-t)/4$, where $t=-\Delta^2$, $\Delta$ being the change in four-momentum in the proton vertex and $Q^2$ is the virtuality of the photon. The $T$ indicates that this calculation is for a transversly polarised virtual photon.

Another approach, widely used for investigations of saturation physics, or the Color Glass Condensate \cite{Gelis:2010nm}, is given by the bSat dipole model \cite{KT, dipole}. Here, the virtual photon splits up into a quark anti quark dipole which subsequently interact with the pomeron and then recombines into a $J/\Psi$ meson. The amplitude for this process is given by:
\begin{eqnarray}
\mathcal{A}_{T,L}^{\gamma^*p\rightarrow\jpsi p}=
i\int_0^\infty 2\pi r{\rm d}r \int_0^1\frac{{\rm d}z}{4\pi}\int_0^\infty 2\pi b{\rm d}b 
(\Psi^*_{\jpsi}\Psi)_{T,L}J_0(b\Delta)J_0([1-z]r\Delta)
\frac{{\rm d}\sigma_{q\bar q}}{{\rm d}^2\vec{b}}
\label{eq:Adipole}
\end{eqnarray}
where $b$ is the impact-parameter, $r$ the size of the dipole, $(\Psi^*_{\jpsi}\Psi)_{T,L}$ is the wave-overlap between the virtual photon and the vector-meson, $z$ is the momentum fraction taken by the quark and $J_0$ is a Bessel function. The dipole cross-section is given by:
\begin{eqnarray}
	\frac{{\rm d}\sigma_{q\bar q}}{{\rm d}^2\vec{b}}=
	2\left(1-\exp\left(-\frac{\pi^2}{2N_c}r^2
	\alpha_s(\mu^2_{\rm dip})xg(x,\mu^2_{\rm dip})T(b)\right)\right)
\end{eqnarray}
where $T(b)$ is the proton thickness in impact-parameter space. It is usually taken to be Gaussian, $T(b)=1/2\pi B_G\exp(-b^2/2B_G)$ where $B_G$ is a parameter which has been fixed by data to be $B_G=4~$GeV$^2$ \cite{dipole}. The scale $\mu^2_{\rm dip}=4/r^2+\mu_0^2$ where $\mu_0^2$ is the starting scale in the DGLAP evolution. Here we follow \cite{dipole} and set $\mu_0^2=1.17~$GeV$^2$. It is easy to see that this dipole cross-section saturates for large $r$ as well as for large gluon densities.

This paper aims at taking the hard scattering limit, where $r\rightarrow 0$, and the non-relativistic limit, where $z=1/2$ in the dipole model eq.\eqref{eq:Adipole}, and show that it becomes equivalent to Ryskin's expression in eq.\eqref{eq:ARyskin} for transveresly polarised photons. In order to do so we will also need to set $\mu^2_{\rm dip}=\mu^2$. 

In section \ref{results} we will show how well Ryskin and the Dipole Model compare when we relax the different approximations.
\section{Taking the hard scattering and non-relativistic limits of the Dipole Model}
The overlap between the wave functions of the transveresly polarised virtual photon and the $\jpsi$ meson is given by \cite{dipole}:
\begin{eqnarray}
	(\Psi^*_{\jpsi}\Psi)_T=\hat{e}_fe\frac{N_c}{\pi z(1-z)}\big( m_f^2 K_0(\epsilon r)\phi_T(r, z)-
	[z^2+(1-z)^2]\epsilon K_1(\epsilon r)\partial_r\phi_T(r, z)\big)
\end{eqnarray}
Where $\hat{e}_f=2/3$ and $m_f$ is the charm quark charge and mass respectively, $e$ is the electron charge, $\epsilon^2=z(1-z)Q^2+m_f^2$ and $K_{0,1}$ are the modified Bessel functions. Here, $\phi_T(r, z)$
is the scalar part of the $\jpsi$ wave-function. It is usually taken to be a "boosted Gaussian" \cite{dipole}:
\begin{eqnarray}
	\phi_{T}=\mathcal{N}_{T}z(1-z)\exp\left(-\frac{m_f^2\mathcal{R}^2}{8z(1-z)}-
		\frac{2z(1-z)r^2}{\mathcal{R}^2}+\frac{m_f^2\mathcal{R}^2}{2}\right)
	\label{eq:boosted}
\end{eqnarray}
where $\mathcal{N}_T$ and $\mathcal{R}$ are parameters which are fixed by data. When we take the 
non-relativistic limit, where $z=1/2$, we instead let the scalar part of the wave function be on the form:
\begin{equation}\label{eq:3}
	\phi_T(r,z)=C\phi_r(r) \delta(z-1/2)
\end{equation}
where $C$ is a constant. When $z=1/2$, we see that eq.\eqref{eq:boosted} becomes a single Gaussian in $r$, and following that we assume the form $\phi_r(r)=\exp(-\beta r^2)$, $\beta$ may be explicitly calculated
through the normalisation conditions on the wave-functions.

The decay width of the vector meson puts a constraint on its wave function. The width is given by:
\begin{equation}
\Gamma_{J/\psi}^{ee}=\dfrac{4\pi \alpha_{em} ^2f_{J/\psi} ^2}{3M_{J/\psi}}
\end{equation}
where the coupling of the vector-meson to the electromagnetic current, $f_{\jpsi}$, is given by \cite{dipole}:
\begin{equation}
f_{J/\psi}=e_f\dfrac{N_c}{2\pi M_{\jpsi}}\int_0 ^1 \dfrac{dz}
{z^2(1-z)^2}(m_f^2-(z^2+(1-z)^2\triangledown_r ^2)\phi_T(r,z))\Big|_{r=0}\\
\label{eq:couplint}
\end{equation}
This fixes the normalisation of the wave-function:
\begin{equation}
C=\sqrt{\dfrac{3M_{\jpsi}^3\pi\Gamma_{J/\psi}^{ee}}{256\alpha^2_{em}e_f^2N_c^2m_f^4}}
\end{equation}
Plugging the scalar wave function into the wave overlap gives in the non-relativistic limit:
\begin{eqnarray}
(\Psi^*_{\jpsi}\Psi)_T&=&
\frac{e_feN_cC}{\pi z(1-z)}e^{-\beta r^2}\delta(z-1/2)\nonumber \\ &~&\big[m_f^2K_0(\epsilon r)+2r\beta(z^2+(1-z)^2)\epsilon K_1(\epsilon r)\big] \nonumber \\
	\label{eq:nonreloverlap}
\end{eqnarray}

For the hard scattering limit we expand the scalar part of the wave-function for small $r$:
\begin{eqnarray}
	\phi_T(r, z)=C\delta(z-1/2)(1-\beta^2 r^2 +\mathcal{O}(r^4))
	\label{eq:hardwave}
\end{eqnarray}
We do the same for the dipole cross-section:
\begin{eqnarray}
	\frac{{\rm d}\sigma_{q\bar q}}{{\rm d}^2\vec{b}}=\frac{\pi^2}{N_c}r^2\alpha_s(\mu^2_{\rm dip})
	xg(x, \mu^2_{\rm dip})T(b)+\mathcal{O}(r^4)
	\label{eq:harddipole}
\end{eqnarray}
We perform the integral over impact-parameter $b$ using the definition of the proton form-factor:
\begin{eqnarray}
\int_0 ^{\infty} 2\pi b{\rm d}b J_0(b\Delta)T(b) \equiv F_N^{2G}(t)	
	\label{eq:formfactor}
\end{eqnarray}
Plugging eqs.\eqref{eq:nonreloverlap}, \eqref{eq:hardwave}, \eqref{eq:harddipole}, and \eqref{eq:formfactor} into eq.\eqref{eq:Adipole}, and performing the integrals over $b$ and $z$ gives: 
\begin{eqnarray}
\mathcal{A}_T^{\gamma^*p\rightarrow J\psi p}&=&
iF_N^{2G}(t)2\pi e_f e C\int_0^\infty {\rm d}r~\alpha_s(\mu^2_{\rm dip})xg(x, \mu^2_{\rm dip})
J_0(r\Delta/2) \nonumber \\ &~&
\left[m_f^2K_0(\epsilon r)+\beta (\epsilon r)K_1(\epsilon r)\right](r^3+\mathcal{O}(r^5))
\label{eq:K1}
\end{eqnarray}
We assume for now that the integral over the second term (containing $K_1(\epsilon r)$) can be neglected (for small $r$, $\epsilon rK_1(\epsilon r)\sim 1$). We will later discuss the impact of this approximation. We further assume that the factorisation and renormalisation scales are independent of $r$, i.e. that $\mu^2_{\rm dip}=\mu^2$ as defined by Ryskin. This assumption will also be discussed later. The integral over the first term then becomes:
\begin{eqnarray}
&~&
\int_0^{\infty}dr  K_0(\epsilon r)J_0(r\Delta/2 )r^3=
	16\frac{2\epsilon^2+t/2}{(2\epsilon^2-t/2)^3} 
\end{eqnarray}
where we have used that $t=-\Delta^2$.

Now the amplitude can be written as:
\begin{eqnarray}
\mathcal{A}_T^{\gamma^*p\rightarrow J\psi p}=
	i4\pi^2\sqrt{\frac{ M_{\jpsi}^3\Gamma_{\jpsi}^{ee}}{3\alpha_{\rm EM}}}
	\alpha(\mu^2)xg(x, \mu^2)F_N^{2G}(t)\frac{2\epsilon^2+t/2}{(2\epsilon^2-t/2)^2}
	\label{eq:finalApprox}
\end{eqnarray}
where we have used that $\alpha_{\rm EM}=e^2/4\pi$. Ryskin makes the approximation that the $\jpsi$ mass is twice that of the charm quark. With this approximation (keeping in mind that $z=1/2$), $\epsilon^2=(Q^2+M_{\jpsi}^2)/4$. Then
eq.\eqref{eq:finalApprox} becomes equal to eq.\eqref{eq:ARyskin}, which is what we wanted to achieve.

\section{Discussion}
\label{results}
\begin{figure}[h]
	\includegraphics[width=0.48\linewidth]{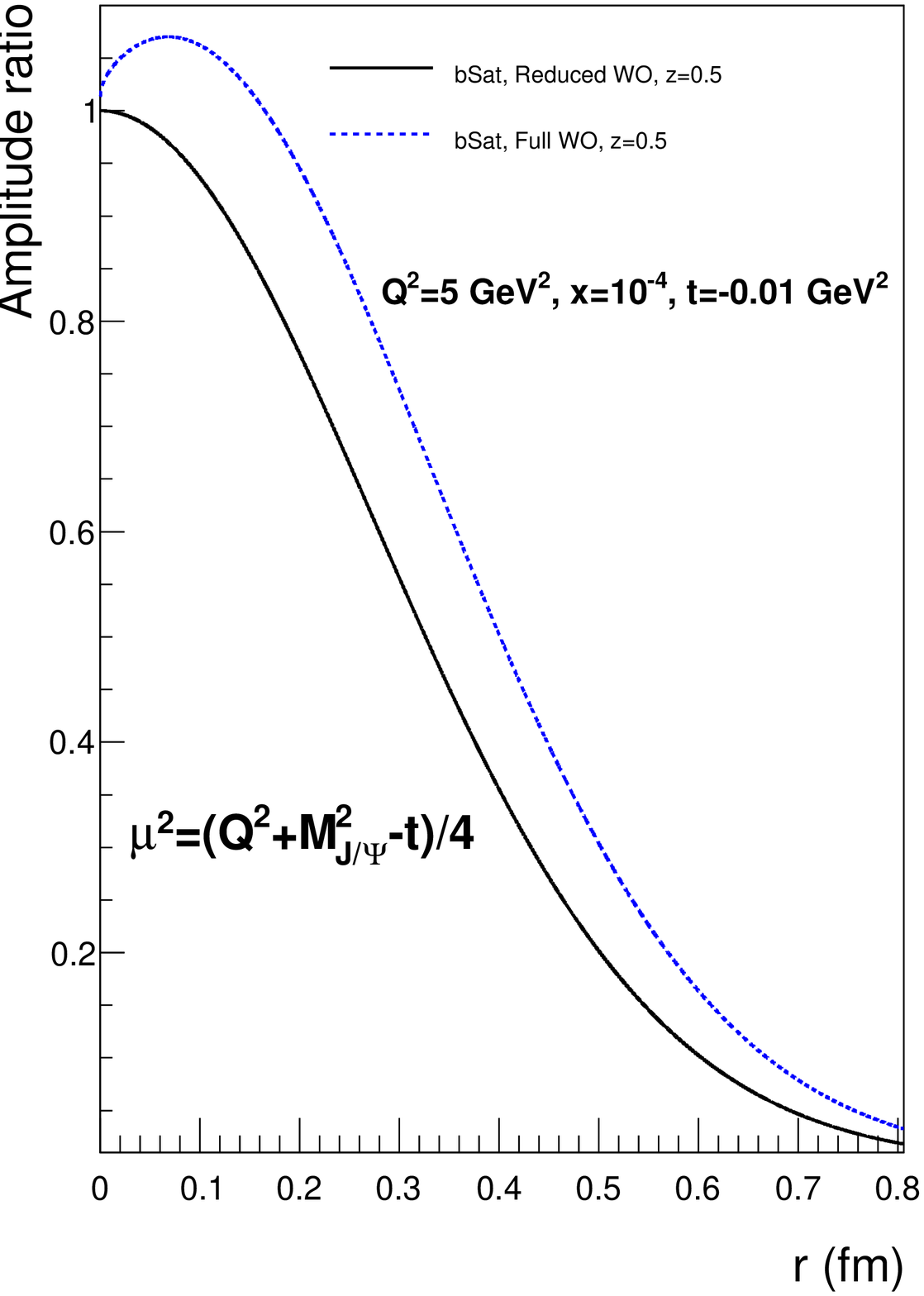}
	\includegraphics[width=0.48\linewidth]{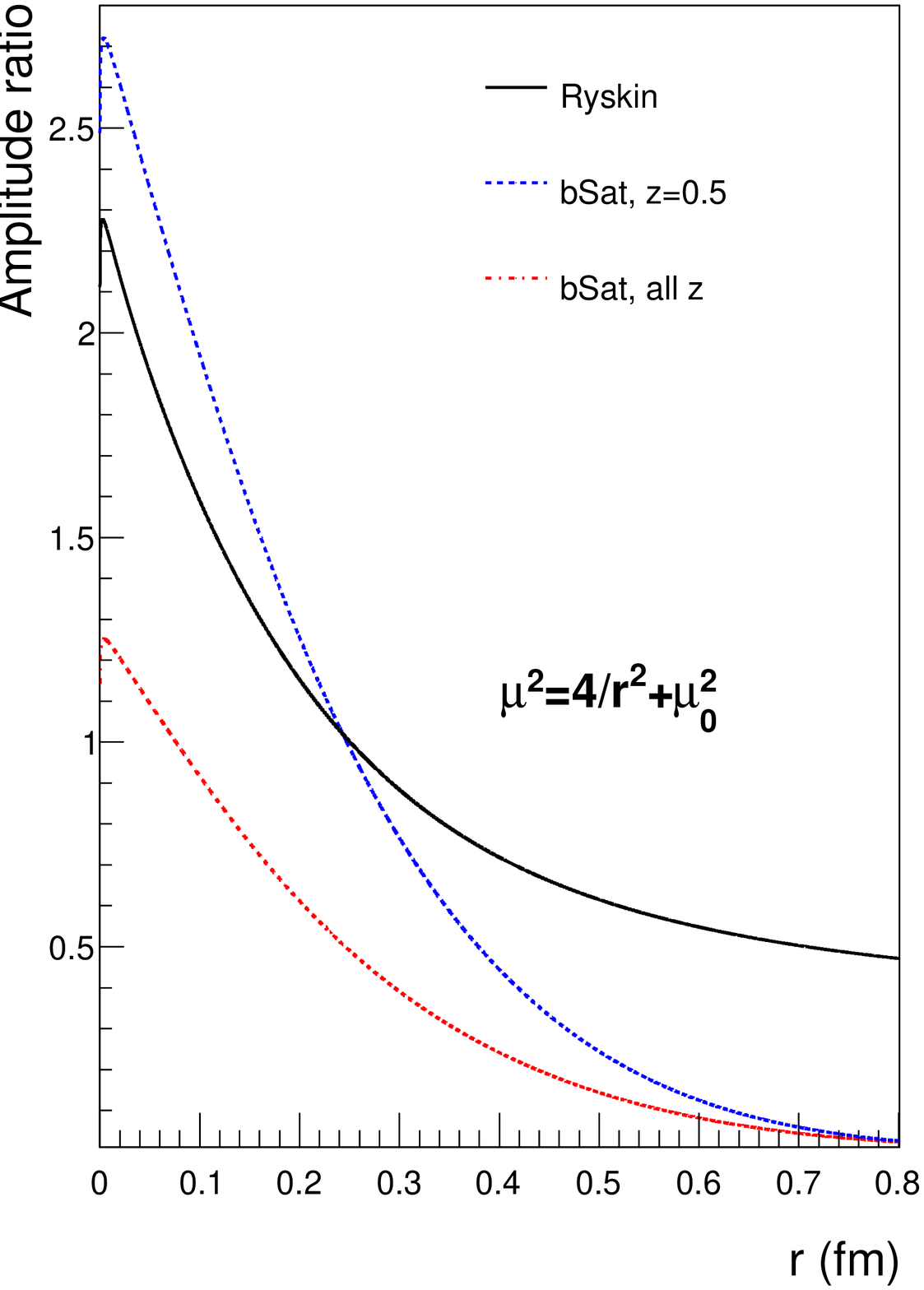}
	\caption{\label{fig:results} 
		Ratios with respect to Ryskin's Amplitude as defined in the text. The left hand side is
		using fixed renormalisation and factorisation scales, while the right plot
		uses scales which are dependent on $r$. All calculation are at $Q^2=5~$GeV$^2$, $x=10^{-4}$, 
		and $t=-0.01~$GeV$^{2}$. All curves are described in detail in the text.}
\end{figure}
In order to analyse the result we define an unintegrated amplitude for Ryskin:
\begin{eqnarray}
	\frac{{\rm d}\mathcal{A}_{\rm Ryskin}^{\gamma^*p\rightarrow J\psi p}}{{\rm d}r}\equiv
	i\sqrt{\frac{\pi^4M_{\jpsi}^3\Gamma_{\jpsi}^{ee}}{48\alpha_{\rm EM}}}
	\alpha(\mu^2)xg(x, \mu^2)F_N^{2G}(t)	K_0(\epsilon r)J_0(r\Delta/2 )r^3
	\label{eq:dAdrRyskin}
\end{eqnarray}
If integrated over $r$ this retains the result in eq.\eqref{eq:finalApprox}. 
The corresponding expression for the Dipole Model is:
\begin{eqnarray}
	\frac{{\rm d}\mathcal{A}_{\rm bSat}^{\gamma^*p\rightarrow\jpsi p}}{{\rm d} r}=
i \pi r \int_0^1{\rm d}z\int_0^\infty  b{\rm d}b 
(\Psi^*_{\jpsi}\Psi)_{T}J_0(b\Delta)J_0([1-z]r\Delta)
\frac{{\rm d}\sigma_{q\bar q}}{{\rm d}^2\vec{b}}
	\label{eq:dAdrDipole}
\end{eqnarray}
where the $\jpsi$ wave function is taken to be the boosted Gaussian in eq.\eqref{eq:boosted}. We take all parameters from \cite{dipole} and fix the decay width in eq.\eqref{eq:dAdrRyskin} accordingly.

In order to compare to the non-relativistic limit we use the following expression:
\begin{eqnarray}
	\frac{{\rm d}^2\mathcal{A}_{\rm bSat}^{\gamma^*p\rightarrow\jpsi p}}{{\rm d} r{\rm d}z}\bigg|_{z=1/2}=
i \frac{r}{2} \int_0^\infty 2\pi b{\rm d}b 
(\Psi^*_{\jpsi}\Psi)_{T,L}J_0(b\Delta)J_0([1-z]r\Delta)
\frac{{\rm d}\sigma_{q\bar q}}{{\rm d}^2\vec{b}}\bigg|_{z=1/2}
	\label{eq:dAdrdzDipole}
\end{eqnarray}
The normalisations of the wave-functions do not exactly match, since the second term in eq.~\eqref{eq:couplint} becomes significant for the boosted Gaussian. We compensate this by multiplying the Ryskin amplitudes by a factor 1.72.

For investigating the effect of using different scales in the strong coupling and gluons density, we simply replace  $\mu^2$ by $\mu^2_{\rm dip}$ in these expressions. We will also investigate the effects from omitting the integral of the second term in eq.\eqref{eq:K1}.

In figure \ref{fig:results} we show ratios with respect to eq.\eqref{eq:dAdrRyskin}. On the left hand side
we show the ratios between eqs.\eqref{eq:dAdrdzDipole} and \eqref{eq:dAdrRyskin} with 
$\mu^2=(Q^2+M_{\jpsi}^2-t)/4$ as in Ryskin. The graph marked "Full WO" uses the entire expression for the wave overlap in eq.\eqref{eq:K1} while the graph marked "Reduced WO" only includes the first term. We see that by including the second term in the wave-overlap we add about 10\% to the amplitude, but for $r\rightarrow 0$, 
the ratio goes to unity, and this omission does not contribute in the hard scattering limit.
The characteristic size of the scattering dipole is given by $1/\epsilon=1/\sqrt{Q^2/4+m_c^2}\sim 0.1~$fm.
We see that here, the hard scattering limit is a good approximation, both ratios are within 10\% of unity.

On the right hand side of figure \ref{fig:results} we show the result of letting the scale in the strong coupling and gluon density have an $r$-dependence in accordance to the dipole model. The graph marked "Ryskin" show the ratio between using $\mu^2_{\rm dip}$ and $\mu^2$ in eq.\eqref{eq:dAdrRyskin}. We see that for small $r$ this increases the amplitude by a factor 2.7. The same is seen in the graph marked "bSat, z=0.5", which is the ratio between eqs.\eqref{eq:dAdrdzDipole} and \eqref{eq:dAdrRyskin} using the dipole scale in the former. However, this effect seems to be somewhat off-set by going away from the non-relativistic limit and allowing all $z$ as is seen in the graph marked "bSat, all z", which shows the ratio between eqs.\eqref{eq:dAdrDipole} and \eqref{eq:dAdrRyskin}. Both "bSat" curves include the full wave-overlap. Around the characteristic size of the dipole, this third ratio is very close to unity which means that Ryskin's predictions for $\jpsi$ production are very similar to those from the dipole model, by the virtue of different approximations off-setting each other.

\section*{Acknowledgement}
This work is S. Anand's undergraduate thesis at Shiv Nadar University. We are very grateful to Tuomas Lappi for suggesting the project and for lending us his comments on its progress. We would like to thank the Physics Department of Shiv Nadar University for its support.

\end{document}